\documentclass[pre,twocolumn,showpacs,byrevtex,showkeys]{revtex4}

\usepackage{amsfonts}
\usepackage{amsmath}
\usepackage{epsfig}

\def\be{\begin{equation}}
\def\eea{\end{eqnarray}}
\def\ee{\end{equation}}
\def\bea{\begin{eqnarray}}
\def\ea{\end{array}}
\def\ba{\begin{array}}

\newcommand{\bel}[1]{\begin{equation}\label{#1}}

\newcommand{\identity}{\small 1\kern-.33em 1}

\providecommand{\openone}{\leavevmode\hbox{\small1\kern-3.8pt\normalsize1}}

\begin{document}
\bibliographystyle{apsrev}

\title{Forecasting extreme events in collective dynamics:\\ an analytic
   signal approach to detecting discrete scale invariance}

\author{G. M. Viswanathan} 

\affiliation{Instituto de Física, Universidade Federal de Alagoas, CEP
57072-970, Macei\'o--AL, Brazil}

\date{\today}

\begin{abstract}{ A challenging problem in 
physics concerns the possibility of forecasting rare but extreme
phenomena such as large earthquakes, financial market crashes, and
material rupture.  A promising line of research involves the early
detection of precursory log-periodic oscillations to help forecast
extreme events in collective phenomena where discrete scale invariance
plays an important role.  Here I investigate two distinct approaches
towards the general problem of how to detect log-periodic oscillations
in arbitrary time series without prior knowledge of the location of
the moveable singularity. I first show that the problem has a definite
solution in Fourier space, however the technique involved requires an
unrealistically large signal to noise ratio.  I then show that the
quadrature signal obtained via analytic continuation onto the
imaginary axis, using the Hilbert transform, necessarily retains the
log-periodicities found in the original signal.  This finding allows
the development of a new method of detecting log-periodic oscillations
that relies on calculation of the instantaneous phase of the analytic
signal.  I illustrate the method by applying it to the well documented
stock market crash of 1987. Finally, I discuss the relevance of these
findings for parametric rather than nonparametric estimation of
critical times.  }
\end{abstract}


\maketitle

\section{Introduction}
\label{sec-intro}

More than a decade of pioneering research involving catastrophic
phenomena as diverse as the rupture~\cite{pre-rupture1} of high
pressure rocket tanks~\cite{rockets}, stock market
crashes~\cite{qf-1999-johansen} and earthquakes~\cite{earthq0} has
lent growing credibility~\cite{physrep,pnas} to the hypothesis that
such extreme events arise due to coherent large-scale collective
behaviors observed in such self-organizing
systems~\cite{pre-rupture1,physrep,pnas,rockets,qf-1999-johansen,earthq0,
earthq1,earthq2,earthq3,earthq4,earthq5,earthq6,earthq7}. An exciting
prospect concerns the possibility of prediction or forecasting of
catastrophic events based on the observation of discrete scale
invariance.  This innovative approach, when applied to problems such
as the prediction of earthquakes or financial crashes, questions the
common assumption that the absence of characteristic scales seen in
self-organizing~\cite{bak-book,pnas} complex systems precludes the possibility
of forecasting~\cite{pnas}.  Instead, prediction becomes possible due
to the appearance of smaller precursory events that in principle can
help determine the critical time $t_c$ of the catastrophic event,
which one can
interpret as a finite time moveable singularity.  One does not
directly observe the singularity due to finite size effects.  Instead
we observe an ultra-large event comparable in magnitude to the system
size.  The best known specific signature of discrete scale invariance
involves log-periodicity~\cite{physrep,pnas,lyra,tsallis,iram-logp}.  Even a
small improvement in the ability to detect log-periodic oscillations
may thus have a relatively large impact, with potentially useful
applications.

Attempts to forecast extreme events by exploiting discrete scale
invariance and log-periodicity implicitly assume an underlying
information-carrying property of some component in the signal studied.
Indeed, in trying to forecast an event for some variable $f(t)$ that
will occur at a future time $t=t_c$, with the
information~\cite{khinchin} available for $f(t)$ at the present time
$t<t_c$, one implicitly assumes the existence of correlations in the
behavior of $f(t)$.  Such correlations imply that the knowledge of the
behavior of $f(t)$ in a certain period necessarily provides
information about the behavior at other (e.g., future) times.  For a
continuous variable $f$, the maximum degree of correlation arises for
holomorphic or analytic $f(t)$, since one can then use analytic
continuation to know $f(t)$ for all future times. No new information
becomes available when time elapses, because $f(t)$ evolves
deterministically.  Consider the well known example of classical
Hamiltonian systems. The fine grained Gibbs entropy, equivalent to the
Shannon information measure, becomes a constant of the motion for such
systems due to Liouville's theorem (see ref.~\cite{zurek} for a
discussion of other information measures).  Indeed, information can
increase only if deterministic evolution becomes interrupted in some
way, e.g., perhaps by some stochastic process.  Such interruptions of
deterministic evolution necessarily lead to breaks in analytic
behavior.  From such considerations, it follows as a logical
consequence that the maximum possible rate of transmission of
information measured in bits per unit time for a physical
communication channel must equal or exceed the rate of occurrence of
non-analytic points in the signal~\cite{fn1}

Detecting log-periodicity presents unique challenges. Direct
parametric estimation to obtain log-periodic fits can fail due to the
presence of extreme fluctuations as well as due to the problem of
large degeneracy of the solutions, i.e. there exist too many good fits
that approximate the best fit.  Parametric methods also fail because
often we do not know which underlying distribution to
assume. Moreover, the large scale catastrophic events of interest may
represent ``outliers'' that do not follow the same distribution as the
smaller scale events.  Hence, most of the research has tended to apply
non-parametric methods.

A widely used spectroscopic method for detecting log-periodic
oscillations involves changing the variable $t$ to a log-time
$\tau\equiv \ln(t_c-t)$, and then studying the power spectrum of the
new series thus generated~\cite{qf-1999-johansen}.  Log-periodic
oscillations will appear periodic in the log-time $\tau.$ However, the
data will no longer appear evenly sampled. Hence standard FFT-based
methods do not work and instead one must obtain the spectrum via the
Lomb periodogram~\cite{numrec}, which can handle unevenly sampled
points. In practice this method works remarkably well.  Since such
nonparametric methods require prior knowledge of the value of $t_c$,
here I investigate the general problem of how we can detect
log-periodic oscillations without having prior knowledge of $t_c$.
Such a method would in principle allow us to ``fine tune'' our
estimates of $t_c$, and then use the nonparametric methods that rely
on {\it a priori} knowledge of $t_c.$ The methods developed here apply
equally to a variety of time series, so a range of applications become
possible.

In this context, one of the most dreaded collective phenomena of our
times relates to the financial and economic crises that have
punctuated our history since the industrial revolution. Economic
events, ranging in size and diversity from the Great Depression to the
ongoing bursting of the real estate bubble in the US, affect an at
least order of magnitude greater number of people than earthquakes or
tidal waves.  Moreover, humans actively participate in the dynamics of
the economy whereas we only passively watch tectonic plate movements.
Indeed, financial crises have the potential to affect almost everybody
(unlike, e.g., earthquakes). For such reasons, this article limits the
application of the new methods developed here to the study of
cooperative economic phenomena. Specifically, I have chosen to focus
on the classic financial ``correction'' of 1987, when stock market
indices dropped $\approx$20\% in an amazing display of cooperative
behavior (i.e., ``herding'') of otherwise rational individual agents.

In Section~\ref{sec-dsi}, I briefly review discrete scale invariance
and log-periodicity. In Section~\ref{sec-method} I develop techniques
for detecting discrete scale invariance.  In Sections~\ref{sec-ex} and
\ref{sec-spc} I illustrate the method and then apply it to the stock
market crash of 1987.  Section~\ref{sec-concl} concludes with a
discussion and a summary.

\section{Complex dimensions and discrete scale invariance}
\label{sec-dsi}

The concept of dimensionality has undergone successive
generalizations: from integer to fractional to negative to
complex~\cite{dimension-evol}.  The fractional and
complex~\cite{earthq5} dimensions have a relation to
fractals~\cite{mandelbrot,bunde-havlin} and scale invariance symmetry,
i.e., when a system's property appears unchanged under a
transformation of scale.  Power laws, such as $f(t)\sim t^\mu,$ play a
fundamental role in describing scale invariance. A change of scale by
a factor $\lambda$ does not alter the power law behavior: $f(\lambda
t)=f(t)\lambda^\mu.$ Real power law exponents usually involve
continuous scale invariance.  In contrast, complex exponents bear a
relation to discrete scale invariance---a discrete rather than
continuous symmetry that holds only for certain discrete values of the
magnification $\lambda_k=\lambda^k$.  The powerful formalism that
emerged from the study and exploitation of scale invariance in
physical systems~\cite{stanley} became known as the renormalization
group (e.g., see ref.~\cite{rg-book}). These advances have led to the
application of fractal concepts and techniques to diverse systems,
ranging from the study of anomalous random
walks~\cite{sergey-pre,facs-luz} and critical points \cite{fulco1999}
to heart dynamics~\cite{heart} and DNA
organization~\cite{dna-bj,dna-pa}.

To model a catastrophic event that corresponds to a moveable
singularity at time $t=t_c$, we can consider an arbitrary signal
$x(t)$ in terms of the renormalization group formalism as follows:
\[
F(t_c-t)\equiv x(t_c) -x(t)
\]
\[
t_c-t'=\phi(t_c-t)
\]
where $\phi$ denotes the flow map and $t_c$ the critical
time~\cite{population,rg-fisher}.  The flow map acts like a ``zoom,''
mapping the time $t$ to a new time $t'$.
  We can then express $F(t)$ as a sum of a singular part and
a non-singular part as follows:
\[
F(t_c-t)= g(t_c-t) +  \frac{1}{\mu} F\left(\phi(t_c-t)\right).
\]
Only the singular part contributes to the ultra-large event, whereas
the non-singular part only describes normal events.  Close to the
critical point, we can apply the linear approximation $\phi(t)=\lambda
t$ to obtain the power law solution, which satisfies
\[ \frac{dF(t_c-t)}{d\ln{[t_c-t]}} = \alpha F(t_c-t) \]
with $\alpha= \ln \mu/\ln\lambda$, i.e., we essentially ignore the
non-singular part.  In practice this solution guarantees that
continuous scale invariance shows up as straight lines on double log
plots, with the slope given by $\alpha$, which plays the role of a
fractal dimension or a scaling exponent. If we allow this dimension or
exponent to become complex, $\alpha=z+ i\omega$, then the power law
$(t_c-t)^\alpha$ becomes $(t_c-t)^z \exp[i\omega \ln[t_c- t]]$, i.e.,
a power law modulated by oscillations with angular frequency $\omega$
in the logarithm of the time---hence the term log-periodic.  Discrete
scale invariance leads to complex exponents $\alpha_n=z+i \omega_n$,
with $\omega_n=2\pi n /\ln \lambda.$

In a number of applications, the first order representation
\be x(t) = A + B(t_c-t)^z +C(t_c-t)^z
\cos[\omega \ln (t_c-t)+\theta ] 
\label{eq-1st-order}
\ee 
captures enough of the relevant behavior to become useful in
forecasting and prediction applications~\cite{qf-1999-johansen}.
Further renormalization group symmetry considerations can lead to
higher order representations useful in some cases~\cite{population}.
A different approach to extending Eq.~\ref{eq-1st-order} involves the
inclusion of higher harmonics. However, the linear approximation leads
us to expect the amplitude of the higher order log-periodic
corrections to decay exponentially fast as a function of the order $l$
of the harmonics~\cite{physrep}. The true behavior (i.e., as opposed
to the linear approximation) of the higher order harmonics leads to a
slower exponential decay of the higher order harmonics. Nevertheless,
the first harmonic still provides a good fit and can account for the
experimental data.  Having reviewed the basics of discrete scale
invariance, I next address the problem of detecting it in arbitrary
time series.

\section{Analytic behavior and discrete scale invariance}
\label{sec-method}

 The relationship between analyticity and information flow discussed
in Section~\ref{sec-intro} has inspired and allowed the development
here of methods for detecting discrete scale invariance in arbitrary
time series.

\subsection{Detecting discrete scale invariance in Fourier space}

\begin{figure*}
\centerline{\psfig{figure=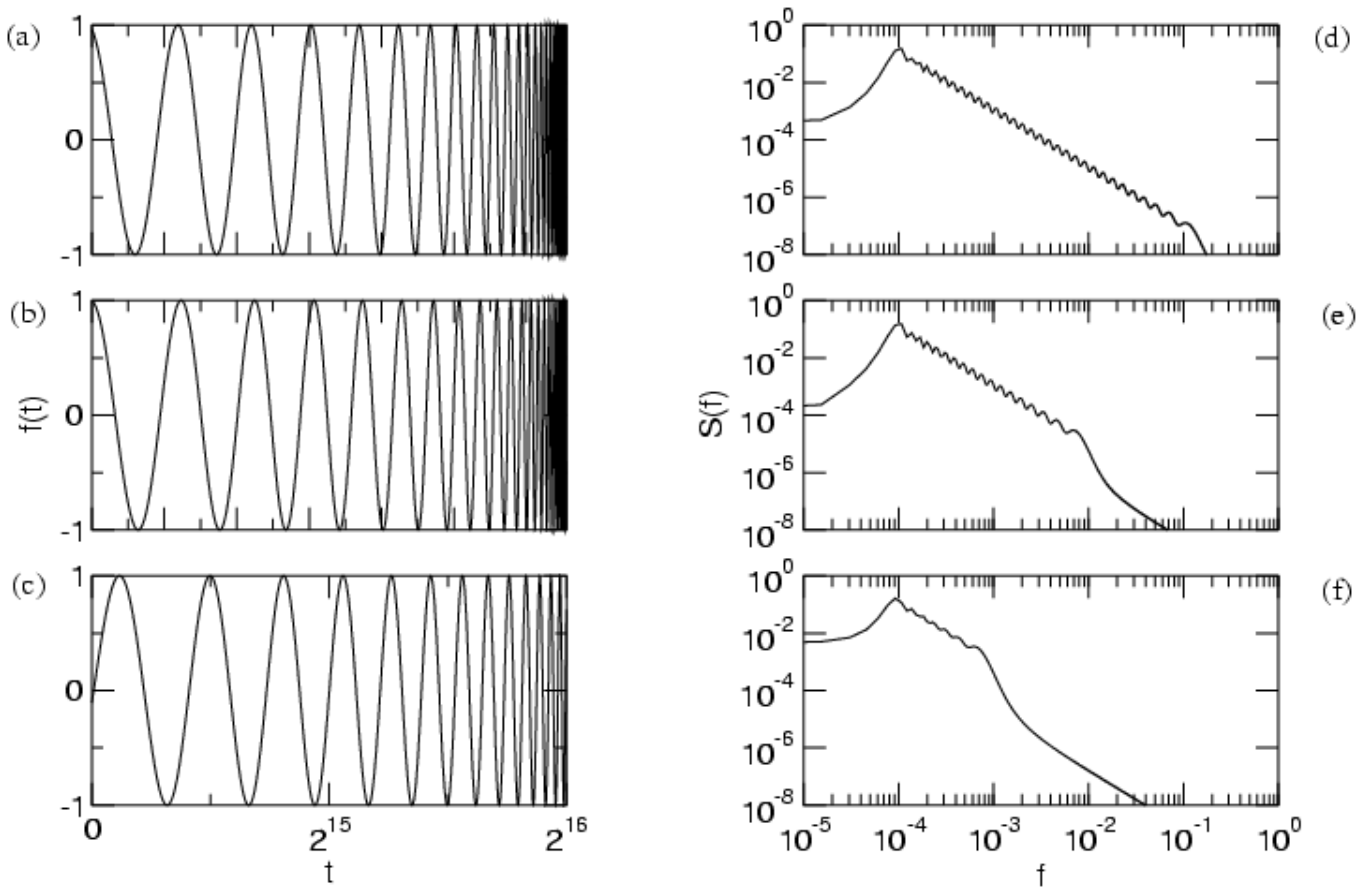,width=18cm,clip}}
\caption{Examples of log-periodic signals
$f(t)=\cos[\omega\ln[t_c-t]]$ shown for $t=1,2,3,\ldots 2^{16}$ with
$\omega=30$ and $t_c=2^{16}+1$ (a), $t_c=65600$ (b) and $t_c=70000$
(c), along with their corresponding power spectra (d,e,f).  An
increase in the critical time $t_c$ leads to a decrease in the upper
cutoff frequency in the log-periodicity of the spectra.  In principle,
one could thus estimate $t_c$.}
\label{fig-example-tc}
\end{figure*}
\bigskip

\begin{figure*}
\centerline{\psfig{figure=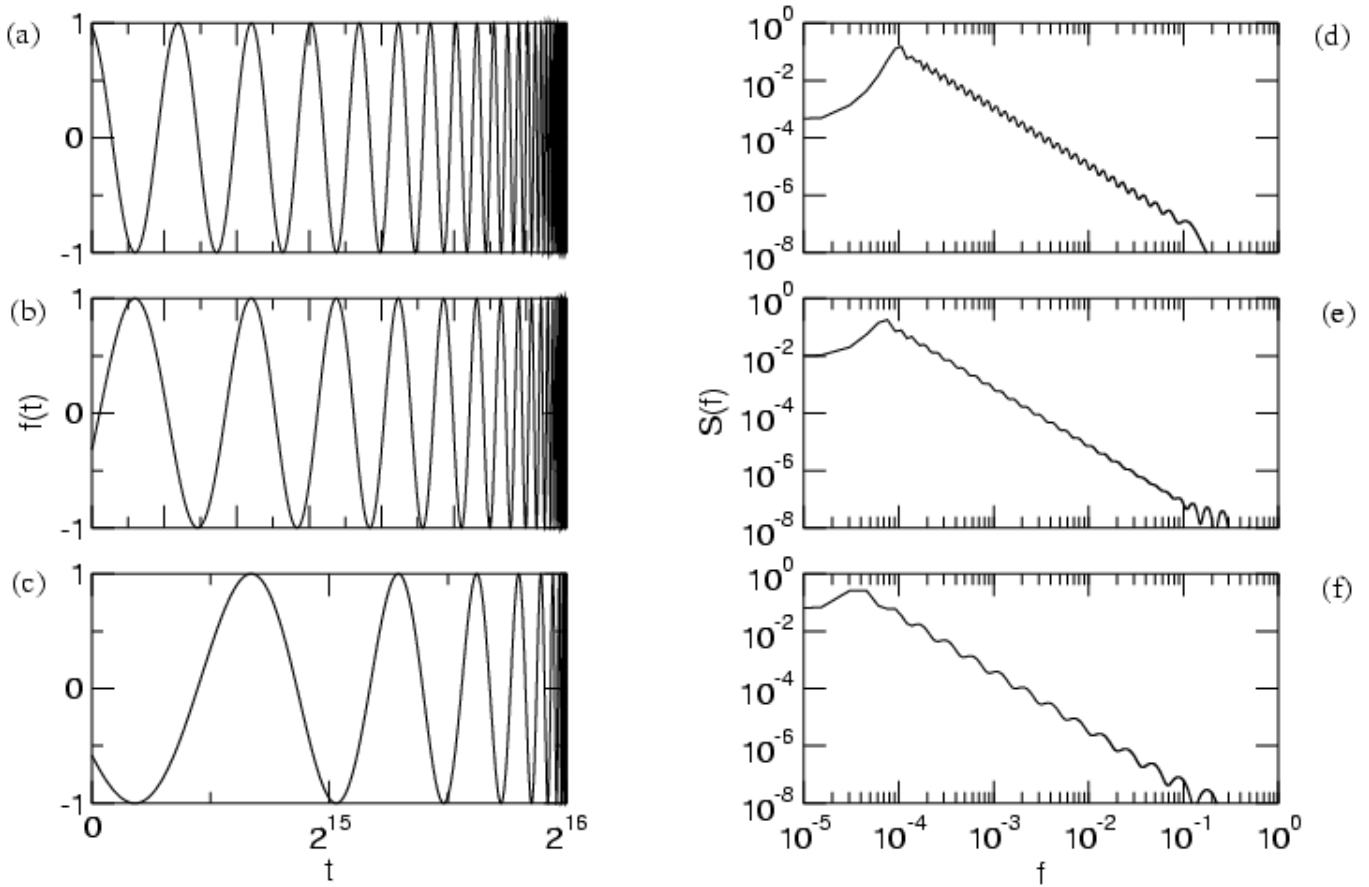,width=18cm,clip}}
\caption{More examples of log-periodic signals
$f(t)=\cos[\omega\ln[t_c-t]]$ shown for $t=1,2,3, \ldots 2^{16} $ with
$t_c=2^{16} +1$ and $\omega=30$ (a), $\omega=20$ (b) and $\omega=10$
(c), along with their respective power spectra (d,e,f).  As a general
rule, for low frequencies at least, discrete scale invariance in the
time domain also manifests itself in the frequency domain.  Notice
that variation of the time-domain angular frequency $\omega$ leads to
a change in the log-periodicity observed in the frequency-domain. In
principle, one could thus estimate $t_c$ knowing $\omega$ and the
upper cutoff frequency in the spectra
(see. Fig~\protect\ref{fig-example-tc}).  In practice, however, we
cannot rely on this method for the more realistic case of noisy data.
This limitation motivates other approaches for estimating $t_c$ for
actual experimentally measured time series.}
\label{fig-example-omega}
\end{figure*}

\begin{figure*}
\centerline{\psfig{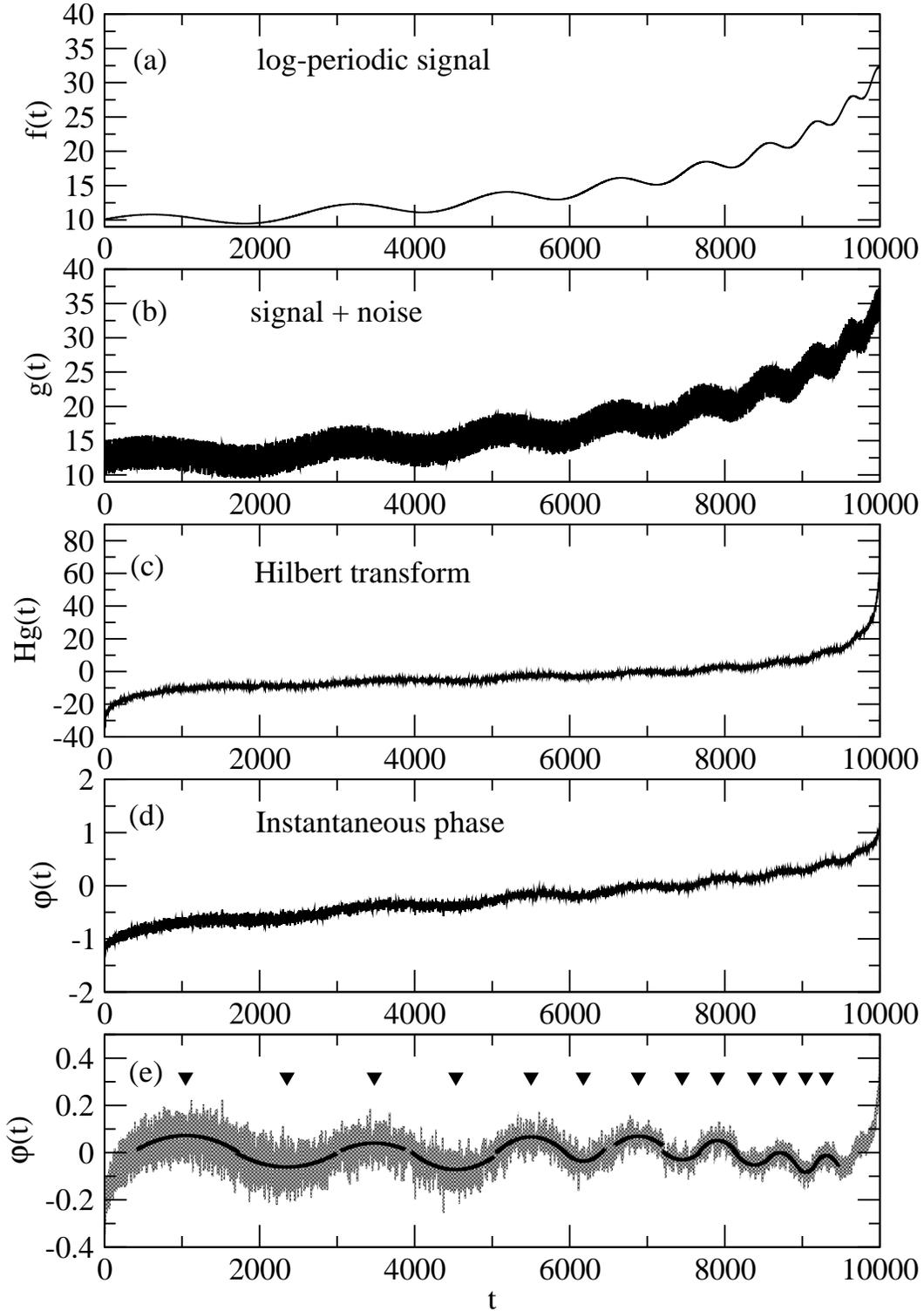}}
\caption{(a) Example of a log-periodicity ``decorating'' a power law,
shown for times $t$ prior to the singularity $t_c$ . Formula used:
$f(t)=\cos(50* \log10(11000-t)) + 1000 * (11000-t)^{-0.5}$.  (b) A
typical noisy time series $g(t)$ containing the log-periodic component
shown in (a) above, with a log-periodic signal-to-noise amplitude
ratio of 2:5. This noisy signal served as test data to illustrate the
method.  (c) Numerically calculated discrete Hilbert transform of the
series shown in (b). (d) Instantaneous phase $\varphi(t)$ of the
analytic signal before and after (e) detrending with polynomial
regression (of order 4). I obtained the positions of the minima and
maxima (shown as triangles) using quadratic regression (parabolic
solid lines) applied in the regions of each peak and valley.}
\label{fig-method}
\end{figure*}

\begin{figure}
\centerline{\psfig{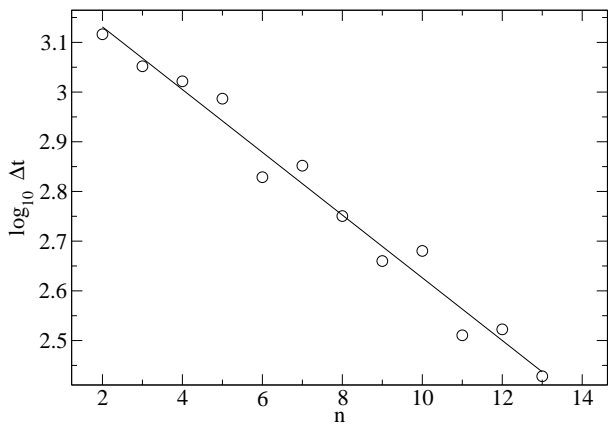}}
\caption{Logarithm of the inter-extrema intervals $\Delta t\equiv
t_n-t_{n+1}$ observed in the log-periodic signal shown in
Fig.~\protect\ref{fig-method}(e) versus the index $n$. The regression
coefficient obtained leads to a value of $\omega= 21.6$. Compare
this with the known value (see Fig.~\protect\ref{fig-method}(a))
$\omega=21.71,$ corresponding to an error of less than 0.5\%.}
\label{fig-times}
\end{figure}

\begin{figure}
\centerline{\psfig{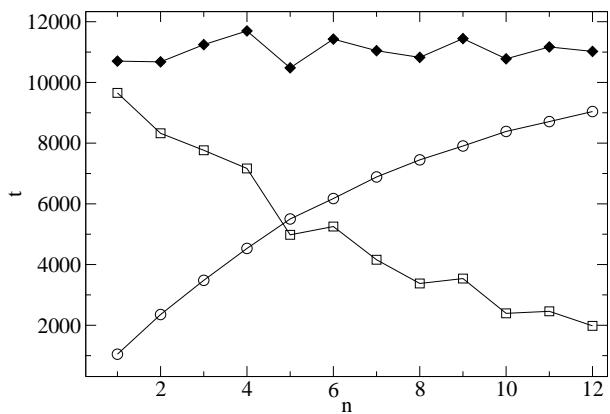}}
\caption{The times (open circles) corresponding to the extrema of the
log-periodicity shown in Fig.~\protect\ref{fig-method}(e), as well as
the product $\Delta t_n/(1-\exp[-\pi/\omega])$ of the inter-extrema
intervals and the scaling factor of the geometric series of
Eq.~\ref{eq-geom} (open squares). Their sum should equal the critical
time $t_c$ and we indeed observe this (filled diamonds). We thus
obtain an estimate $t_c=11040\pm 370$. Compare this estimate with the
known value $t_c=11000.$ The discrepancy lies below 1\% and the
standard deviation represents less than 5\%.}
\label{fig-tc}
\end{figure}

\begin{figure}
\centerline{\psfig{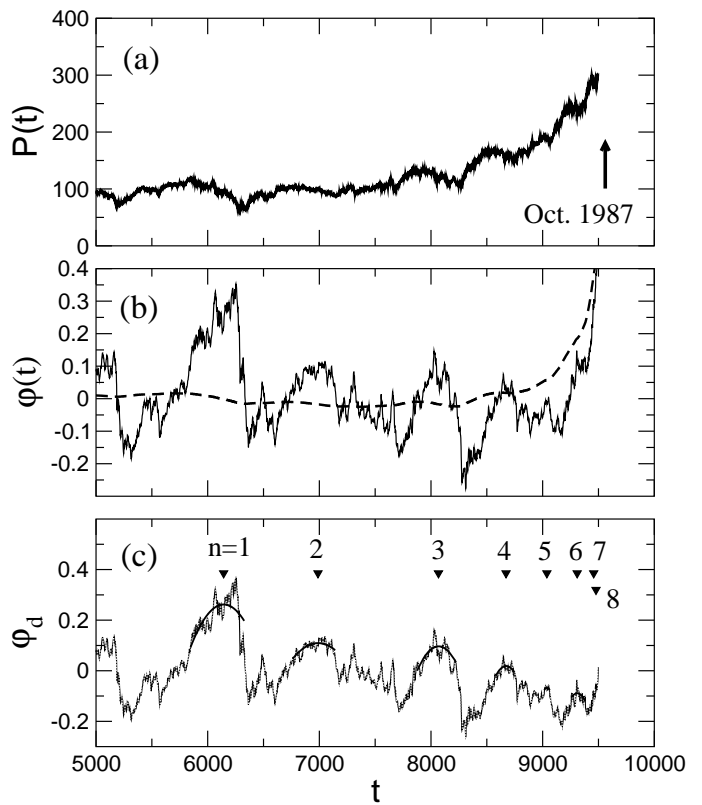}}
\caption{(a) Price $P(t)$ corresponding to the S\&P index shown for
approximately 5000 business days, culminating in the ``great
correction'' of October 1987. The forecasting tests used only the data
shown in black and the grey data near the crash appears solely for
visual clarity (with time $t=0$ chosen arbitrarily).  (b) the
instantaneous phase $\varphi(t)$ after polynomial regression and (c)
after further detrending by subtraction of the cumulative average
(dashed line) defined in Eq.~\ref{eq-ma}, resulting in a detrended
phase signal $\varphi_d(t)$.  The estimated positions of the maxima
come from quadratic regression.  I have indicated their positions
labelled by the index $n.$ Notice that the peaks at $n=7$ and $n=8$
appear nearly indistinguishable, but they actually appear distinct at
smaller scales (not shown). The decreasing intervals for successive
$n\geq 2 $ suggests the possibility of discrete scale invariance. }
\label{fig-spc}
\end{figure}

\begin{figure}
\centerline{\psfig{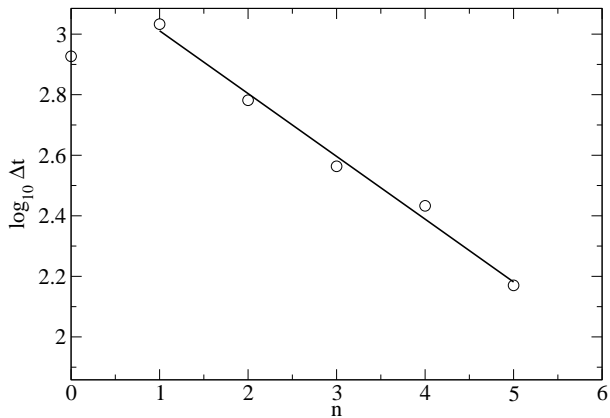}}
\caption{ Logarithm of the inter-maxima intervals $\Delta t\equiv
t_n-t_{n+1}$ observed in the signal shown in
Fig.~\protect\ref{fig-spc} versus the index $n$. The regression
coefficient obtained leads to an 
estimate of $\omega$ (see
Table~\protect\ref{tab-tc})}
\label{fig-spctimes}
\end{figure}

\begin{figure}
\centerline{\psfig{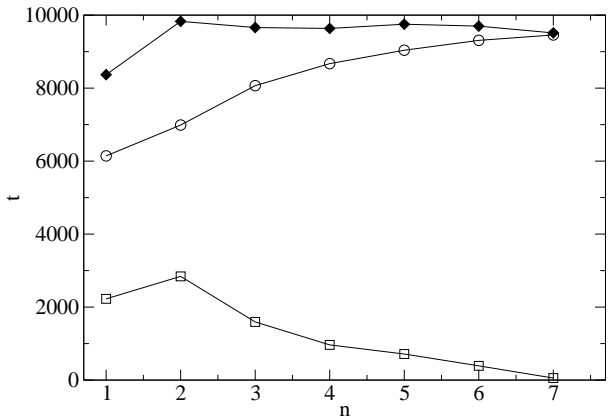}}
\caption{The times (open cicles) corresponding to the maxima of the
log-periodicity shown in Fig.~\protect\ref{fig-spc}(c), as well as the
product $\Delta t_n/(1-\exp[-\pi/\omega])$ of the inter-maxima
intervals  and the scaling factor of the geometric
series of Eq.~\ref{eq-geom-spc} (open squares).
We thus obtain an estimate for $t_c$ (filled diamonds).  
Table~\protect\ref{tab-tc}
 summarizes
the results.}
\label{fig-spc-tc}
\end{figure}

How can we exploit the role of correlations to detect log-periodicity
without prior knowledge of $t_c$?  The power spectrum, defined as the
modulus squared of the Fourier transform of the time series, allows us
to measure correlations. Indeed, one could also equivalently define
the power spectrum as the Fourier transform of the standard two-point
autocorrelation function.  As a starting point, let us consider a
useful but not widely known fact about log-periodic time series and
their Fourier transforms.  Discrete scale invariance in a time series
can sometimes also become manifest in the frequency domain.

Mathematically, log-periodicity in the time domain appears in the
frequency domain because complex exponents also appear in the Fourier
transform of the signal.  Consider the indefinite Fourier integral
\[\begin{split} \int & {\rm d} t \exp[i \omega' t ] 
\cos\big[ \omega \ln[t_c-t]\big] \\ &
=\displaystyle\frac{1}{2\omega'}
\biggl[ (t_c-t)^{-i\omega}[i\omega'(t_c-t)]^{-i\omega} \\ &
\phantom{=\displaystyle\frac{1}{2\omega'} 
\biggl[} \bigl(
(i\omega'(t_c-t))^{2i\omega} ~\Gamma[1-i\omega,i\omega'(t_c-t)] \\ &
\phantom{=\displaystyle\frac{1}{2\omega'} \biggl[} 
+(t_c-t)^{2i\omega}
~\Gamma[1+i\omega,i\omega'(t_c-t)] \bigr) \\ &
\phantom{=\displaystyle\frac{1}{2\omega'} \biggl[} 
(-i\cos[\omega'
t_c] + \sin[\omega' t_c]) \biggr] \; \; ,
\end{split} \;\; 
\]
where $\Gamma[\cdot,\cdot]$ denotes the upper incomplete Gamma
function.  This algebraically defined indefinite integral deals with
complex quantities and represents an antiderivative with respect to
the integration variable $t$.  Note the terms of the form ~$\omega' \;
^{-i\omega}$.  In practice we can evaluate this integral with lower
and upper integration limits $t=-\infty$ and $t=t_c-\epsilon$ to
calculate the Fourier transform.
Figs.~\ref{fig-example-tc} and \ref{fig-example-omega} show
log-periodic time series and their power spectra $S(f)$, defined as
the modulus squared of the Fourier transform of the time series. We
clearly see that the log-periodicity in the time domain manifests
itself as log-periodicity in the frequency domain, within a range of
frequencies.  Specifically, the log-periodic scaling breaks down in
the spectra at low and high frequencies and the cutoff frequencies
depend both on the value of $\omega$ as well as the temporal
separation from the singularity.  Except for these high and low
frequency cutoffs, discrete scale invariance in the time domain
manifests itself as discrete scale invariance in the frequency domain.
Note that a similar relation holds for continuous scale invariance,
i.e. the Fourier transform of a power law tailed function $f(t)\sim
t^{-\alpha}$ can also have a power law behavior (e.g., at low
frequencies).

An increase in the critical time $t_c$ leads to a decrease in the
upper cutoff frequency in the log-periodicity of the spectra. In
principle one could exploit this relationship. Indeed, we can show in a
straightforward manner that discrete scale invariance in Fourier space
will break down near an upper cutoff frequency $f_{\rm high}$ given by
\bel{eq-cutoff}
\ln [1+ 1/(f_{\rm high}(t_c-t_{\rm max}))]\approx 2\pi/\omega\;\; ,
\ee
where $t_{\rm max}$ denotes the largest time contained in the time
series. 
Except for too small $\omega$, we can approximate
\bel{eq-cutoff2}
f_{\rm high}\approx \frac{\omega}{2\pi (t_c-t_{\rm max})}\;\; .
\ee
Hence, one could thus estimate $t_c$ knowing $\omega$ and the upper
cutoff frequency in the spectra. We find that this relationship agrees
fairly well for the data shown in Fig.~\ref{fig-example-tc}.  Yet, in
practice I expect that this method may not work very well with real
data.  For realistic time series, the spectrum contains too many other
features that drown out the log-periodic behavior.  Perhaps one cannot
systematically and reliably apply this method to detect log-periodic
oscillations in realistic scenarios.  Nevertheless, the concept
appears to have validity on a fundamental level.

Moreover, these findings offer insight about the potential of
investigating analytic behavior for detecting the crucial
log-periodicities.  Indeed, the basic premise of forecasting based on
exploiting ``hidden'' analytic properties appears valid.  Continuing
this line of reasoning leads to a second approach.
Next, 
I show below that the analytic signal obtained using the Hilbert
transform of a time series can help to isolate the log-periodic
signature in arbitrary time series.  
I briefly outline the method here before describing it detail below.
The method involves taking the Hilbert transform of the time series
to obtain the quadrature or analytic signal, consisting of the
instantaneous amplitude and the instantaneous phase or argument. If
the time series behaves purely log-periodically $\sim\exp[i \omega \ln
(t_c-t)]$, then the phase will behave logarithmically, $\sim\omega \ln
(t_c-t).$ On the other hand, experimentally obtained data will
realistically always have non-log-periodic components, so a pure
log-periodic series represents an impractical idealization.

In the more realistic case of a time series with a small log-periodic
component, the analytic signal phasor will rotate log-periodically in
the complex plane not around the origin, but rather around some other
point on the complex plane. In principle this ``center'' or ``focus''
of log-periodic rotation can itself fluctuate in time, due to the many
other components that contribute to the analytic signal.  Hence, in
the case of a small log-periodic component in the time series, the
phase of the analytic signal will not vary as $\sim \omega \ln
(t_c-t)$. Instead the phase will have a component that oscillates
log-periodically, due to the angle subtended at the origin by the
analytic signal phasor tip.  The log-periodicity contained in a time
series need not necessarily appear in the amplitude of the analytic
signal (e.g., consider how a pure log-periodic oscillation has
constant amplitude). In contrast, the log-periodicity necessarily
appears in the phase of the analytic signal. By studying the
instantaneous phase, we may thus enhance or highlight the
log-periodicity apparent in time series.

\subsection{The Hilbert transform}
\label{subsec-ht}

The Hilbert transform $Hf(x)$ of a function $f(x)$ represents the
convolution of the function $f$ with $1/\pi x$.
Mathematically, we define the Hilbert transform as a Cauchy principal
value,
\[ Hf(x) \equiv \lim_{\epsilon\rightarrow 0, R\rightarrow \infty}
\left[
\int_{-R}^{x-\epsilon}   dy\frac{f(y)}{y-x}+
\int^{-R}_{x+\epsilon}   dy\frac{f(y)}{y-x} 
\right] \;,
\]
to avoid the singularity at $x$. Moreover, cancellation towards $\pm
\infty $ allows non-integrable functions to have well defined Hilbert
transforms. The Hilbert transform also corresponds to the inverse
Fourier transform of the product of the Fourier transform of $f(x)$
with $i \mbox{~sgn}(x)$ (where the latter gives the Fourier transform
of $1/\pi x$). Hence the Hilbert transform effectively maintains the
Fourier amplitudes but shifts all phases by $-\pi/2$. 
Hence
$H^2=-\openone.$

The Hilbert transform of $x(t)$ allows us to define an analytic signal
\[
\chi(t)\equiv x(t) + iHx(t) = A(t) \exp[i\varphi(t) ]\;,
\]
where $A(t)$ and $\varphi(t)$ represent the instantaneous amplitude
and phase of the signal.  This and related properties have led Hilbert
transforms to have important and diverse applications (e.g., see
refs.~\cite{hilbert-ieee,hilbert-raul,hilbert-sornette}).

\subsection{Analytic signal of log-periodic data}
\label{subsec-as}

For a pure log-periodic analytic signal
\[\chi(t)=\exp[i \omega\ln[t_c-t]]\]
the ``unwrapped'' instantaneous phase will follow
\[\varphi(t)=\omega\ln[t_c-t].\] Since in Nature we typically observe
log-periodicity ``decorating'' a power law or else hidden in noisy
data, we must consider an analytic signal of the form
\[\chi(t)=\chi_0(t)+ a\exp[i \omega
\ln[t_c-t]=A(t)\exp[i\varphi(t)] \;\; ,
\] 
with $a\ll \chi_0(t)$. Let
$\chi_0(t)=A_0(t)\exp[i\varphi_0(t)]$. Then
\[ A(t)=\sqrt{A_0^2(t)+a^2+2A_0(t)\; a \cos\big[\varphi_0(t)-\omega\ln[t_c-t]\big]}
\]
 and
\[\varphi(t)=
\arctan\left[\frac{A_0(t)\sin[\varphi_0(t)]+a\sin\big[\omega\ln[t_c-t]\big]}
{A_0(t)\cos[\varphi_0(t)]+a\cos\big[\omega\ln[t_c-t]\big]}\right] \;\; .
\]

Notice that log-periodicity need not appear in $A(t)$, by considering
for instance, the important case $A_0(t)=0$, for which we obtain
$A(t)=a=$ constant.  In contrast, log-periodicity will always appear
in the phase $\varphi(t)$, which contains the information arising from
the complex exponents associated with discrete scale invariance.  The
phase, calculated as an $\arctan$ will belong to a single branch on
the complex plane, but we can ``unwrap'' the phase to make it vary
outside the conventional range $-\pi/2 \leq \varphi \leq \pi/2,$ to
avoid abrupt discontinuities at the branch cut.

\section{An illustrative example}
\label{sec-ex}

The technique developed above finds practical application to arbitrary
time series. Consider, as an illustrative example, the log-periodic
oscillation decorating a power law shown in Fig.~\ref{fig-method}(a).
In a realistic situation, we would never observe such a clean
signal. Rather, it would be embedded in noise or added to other types
of signals.  Fig.~\ref{fig-method}(b) shows the same log-periodic
signal with noise added. Our goal involves finding the critical time
from such data. 

Let us thus use the signal shown in Fig.~\ref{fig-method}(b) as our
test data. Applying the method described in the previous section,
Fig.~\ref{fig-method}(c) shows the discrete Hilbert transform of the
signal and Fig.~\ref{fig-method}(d),(e) show the instantaneous phase
calculated from the analytic signal before and after detrending.

The last plot, shown in Fig.~\ref{fig-method}(e), permits an
estimation of the times of the minima and maxima. I have used
quadratic regression fits in the region of the extrema, due to the
validity of the parabolic approximation. In principle, other methods
may work equally well.  Alternatively, one could estimate the times
corresponding to the zeroes rather than the extrema. Yet another
possibility includes direct parametric fitting of a log-periodic
cosine function. I have not used this direct parametric estimation due
to the reasons mentioned in Section~\ref{sec-dsi}.

Nevertheless, knowledge of the times of the extrema permits parametric
estimation of the log-periodic angular frequency $\omega$
Fig.~\ref{fig-times}. I have used
\bel{eq-delta-t} \Delta t_n = \exp[-\pi n/\omega](\exp[-\pi
\omega]-1) \;\; , \ee
where $n$ denotes an arbitrary integer index to identify successive
extrema and $\Delta t^{(n)}\equiv t_{n+1}-t_n$ represent the
inter-extrema intervals.  The limit $n\rightarrow \infty$ leads to the
singularity.  The regression coefficient obtained leads to a value of
$\omega= 21.6$, compared with the known value $\omega=21.71.$
Notice the remarkable agreement.

Once we have knowledge of $\omega$ and the positions of the extrema
(or zeroes), it becomes straightforward to find the critical time.
From Eq.~(\ref{eq-delta-t}) it follows that the inter-extrema intervals
follow a geometric series defined by 
\bel{eq-geom}
\frac
{\Delta t_{n+1}}
{\Delta t_n}
=
 \exp[-\pi/\omega] \;\; .
\ee 
The critical time $t_c$ thus satisfies
\bel{eq-tc}
t_c=t_n + \frac{\Delta t_n}{1-\exp[-\pi/\omega]} \;\; .
\ee
Fig.~\ref{fig-tc} shows the times $t_n$, the inter-extrema intervals
$\Delta t_n$, and the estimated values for $t_c$. We obtain an
estimate for $t_c$ in excellent agreement with the known value.

Previous applications of the Hilbert transform to study log-periodic
precursors (e.g., of financial crashes~\cite{hilbert-sornette}) have
assumed prior knowledge of $t_c$.  The
log-time~\cite{hilbert-sornette} $\tau\equiv \ln(t_c-t)$ behaves not
log-periodically, but periodically, rendering the use of Hilbert
transforms useful. In contrast, here we have assumed no prior
knowledge of $t_c$.  Rather, such knowlegde constitutes the goal.

In summary, the method uses the following steps: (i) generation of the
analytic signal from the original time series, (ii) extraction of the
instantaneous phase and any necessary ``unwrapping'' of the phase,
(iii) detrending with polynomial regression etc., (iv) testing for
evidence of log-periodicity using more conventional methods.  If
applicable, then the final and most important step consists of
estimating the location of the moveable time singularity $t_c$ by
regression methods.

The above example illustrates step by step the application of the
method to arbitrary time series.  Below I apply the method to actual
empirical data.  Indeed, the crucial question concerns whether the
method works for actual experimental data.  I have chosen for this
purpose the stock market crash of 1987, since it represents a well
known event in which the collective social behavior of individual
economic agents unleashed financial havoc and in which systematic
studies have documented the role played by log-periodicities.

\section{The stock market crash of 1987}
\label{sec-spc}

\begin{table*}[t]

\begin{center}
\begin{tabular}{| c | c | c | c |}
\hline
Data points from Fig.~\protect\ref{fig-spctimes}  & $\omega$ & $t_c$ & Calendar dates \\
\hline
$n=1,8$ excluded & 0.076 & 9681 $\pm$ 107 &  2/10/1987---8/8/1988 \\
$n=1$ excluded & 0.110 & 9419 $\pm$ 179 & 9/6/1986---5/11/1987 \\
All points & 0.090 & 9548 $\pm$ 86  & 24/4/1987---29/12/1987 \\
\hline
\end{tabular}
\end{center}
\bigskip
\bigskip
\bigskip
\caption{Estimates of the log-periodic angular frequency $\omega$ and
  $t_c$ found from Fig.~\protect\ref{fig-spctimes}. The first and last
  points in Fig.~\protect\ref{fig-spctimes} ($n=1,8$) deviate further
  than the middle points from an exponential behavior. Therefore, I
  have shown three estimates for $\omega$: (i) taking all points into
  account, (ii) taking only the middle points into account, excluding
  the first and last points, and (iii) taking all but the the first
  point into account.  (The {\em a priori} known value is $t_c\approx
  9582.$) In all three cases, I dropped the $n=1$ point to estimate
  $t_c$, since it clearly does not belong to the same regime as the
  other maxima, as seen in Fig.~\protect\ref{fig-spc-tc}.  Remarkably,
  all three estimates for $t_c$ surprisingly bear consistency with the
  actual stock market crash that followed in the middle of October
  1987.}

\label{tab-tc}

\end{table*}

Fig.~\ref{fig-spc}(a) shows data corresponding to approximately 5000
business days of the S\&P500 financial index~\cite{econo1} (which has
ticker symbol ``SPC''). The area in grey shows the crash of October
1987, when stock markets lost some 20\% of their valuation.  Since the
idea behind the proposed method involves forecasting the crash, I
applied the method only to the data shown in black. The data in grey
appears only for greater visual clarity.

Fig.~\ref{fig-spc}(b) shows the instantaneous phase $\varphi'(t)$
after detrending with polynomial regression.  To emphasize the
oscillations, I have further detrended the data by subtracting a
uniformly weighted moving average $y(t)$ with varying window size,
using
\bel{eq-ma}
y(t)= \frac{1}{t_{\rm max}-t+1}\sum_{t'=t}^{t_{\rm max}} \varphi'(t') \;\; , 
\ee
where $t_{\rm max}$ corresponds to the last point of the series.  I
arbitrarily have chosen 17 June 1987 as the last data point included
in the test. This date corresponds to some 80 business days (i.e.,
several months) antecedent to the crash of October 1987.

Fig.~\ref{fig-spc}(c) shows the phase after detrending with $y(t)$.  I
have estimated the positions of the maxima using using parabolic
regression and have arbitrarily labelled each maxima with an index
$n=1,2,.\ldots, 8$.  Notice that for $n\leq 2$ no sign appears of the
characteristic log-periodicity. However, for $n\geq2$ the intervals
between maxima become smaller, consistent with a possible
log-periodicity.  I have chosen to show only the maxima, because the
minima appear less clear. So in what follows, note that I use a phase
variation of $2\pi$ for successive $n$, in contrast to the a phase
variation of $\pi$ for the inter-extrema intervals. So we must replace
Eq.~\ref{eq-geom} by 
\bel{eq-geom-spc}
\frac
{\Delta t_{n+1}}
{\Delta t_n}
=
 \exp[-2\pi/\omega] \;\; ,
\ee
and Eq.~\ref{eq-tc} similarly becomes
\bel{eq-spc-tc}
t_c=t_n + \frac{\Delta t_n}{1-\exp[-\pi/\omega]} \;\; .
\ee

Fig.~\ref{fig-spctimes} shows the inter-maxima intervals $ \Delta
t_n$. Notice the approximate exponential behavior.  Using regression,
we can estimate the log-periodic angular frequency $\omega_1$. Then, we
can apply Eq.~\ref{eq-tc} to estimate the crash, assuming that it
should occur near the singularity. Fig.~\ref{fig-spc-tc} shows the
estimates for $t_c$ and Table~\ref{tab-tc} summarizes the results.  All
estimates contain the actual crash within the margins of
error. Clearly, we must reject any estimate for $t_c$ that occurs
within the dates studied, since we know {\it a priori} that no such
event transpired.

\section{Concluding remarks}
\label{sec-concl}

Several points deserve commenting.  One thought-provoking point
concerns the analysis of the 1987 crash.  The results reported here
might bear some relation to previous findings. The very large-scale
log-periodic oscillations seen in Fig.~\ref{fig-spc} are not
inconsistent with similar conclusions in previous studies~\cite{pnas}.
Such results raise a number of questions and the implications merit
further study. Was the financial crisis of 1987, which few had
anticipated even one month prior to the crash, really being slowly
built up over large time scales spanning years? Were the individual
and institutional agents really starting to behave collectively so
long before the crash? What about the implication for the individual
agents: is our apparently ``free will'' essentially irrelevant to
collective dynamics?

Moreover, in the past few years, financial institutions (and some
individuals) have started to use automated trading software to buy and
sell financial assets (in real time~\cite{econo2}). Each algorithmic
``robo-trader'' \cite{robotrader} follows a given set of rules of
arbitrary complexity, however other traders (human and robotic) do not
know the specific rules and thus cannot exploit them to obtain
financial arbitrage.  How will the advent of large numbers of
robo-traders affect the collective dynamics and what are the
implications relating to the probability of financial crises?

Another noteworthy aspect concerns the the general nature of the
methods developed here, whose application goes beyond financial data.
One of the major difficulties in parametric estimation of log-periodic
properties is due to the lack of foreknowledge of $t_c$. It is not
inconceivable that even small improvements in the ability to estimate
$t_c$ can be of potential use in forecasting research.  It would be
interesting to apply and further study the method using other data
sets.  Given the recent application of physical methodologies to study
music time series~\cite{music}, it is even conceivable that such
methods could be applied to obtain quantitative descriptions of
aesthetic phenomena in the arts.  In fact, wherever cooperative
effects are involved, there is a real possibility that discrete scale
invariance plays some role.  For example, an interesting question, in
this context, is whether the method is applicable to coherent noise
phenomena~\cite{cn1,cn2,cn3}.  Concerning the method itself, there is
room for further improvement, e.g., corrections for finite size
effects of the conventionally used Hilbert transform algorithm.
Similarly, one could take into consideration the role of log-periodic
harmonics, which no doubt play an important role in such phenomena.
Such issues are important but their relevance is secondary to the more
significant question of the theoretical basis for the method. The
inclusion here of further discussion about such secondary issues would
detract from the central focus.

In summary, I have investigated two distinct approaches towards the
general problem of how to detect log-periodic oscillations in
arbitrary time series without prior knowledge of the location of
critical time.  The more promising method involves analytic
continuation of the signal onto the imaginary axis, using the Hilbert
transform. I have shown that the instantaneous phase necessarily
retains the log-periodicities found in the original signal and develop
a new method of detecting log-periodic oscillations.  Initial results
of the application of the method to the stock market crash of 1987
motivate further systematic studies to verify how much promise this
approach holds for forecasting extreme events.

\section*{Acknowledgements}

I thank FAPEAL, CAPES and CNPq for financial support and 
I. M. Gleria,
F. A. B. F. de Moura,
J. M. Hickmann, M. G. E. da Luz, M. L. Lyra, R.
Montagne and D. Sornette for comments.

\bibliography{citedworks}

\begin{thebibliography}{44}
\bibitem[{fn1()}]{fn1}
\bibinfo{note}{As an unusual but entertaining exception, consider the anomalous
  pulse given by \mbox{$f(t)\propto 1-\exp[-1/t^2]$} which has vanishing
  derivatives and peculiar properties at $t=0$.}
\end{thebibliography}

\end{document}